\documentclass[prb, aps, twocolumn, superscriptaddress]{revtex4}

\usepackage{graphicx, epsfig}
\usepackage{amssymb,amsmath}

\newcommand{\pn}[1]{}
        \def \k{\kappa}
       \def \th{\theta}
    \def \k{\kappa}

\def\lba{\left(}    \def\rba{\right)}
\def\lbc{\left[}    \def\rbc{\right]}

\newcommand{\bra}[1]{\langle\left.{#1}\right|}
\newcommand{\ket}[1]{\left|{#1}\right.\rangle}
\newcommand{\xpct}[1]{\langle{#1}\rangle}    

\DeclareMathOperator{\tr}{tr}

\def \be{\begin{equation}}
\def \ee{\end{equation}}
\def\bea{\begin{eqnarray}}
\def\eea{\end{eqnarray}}

\def\be{\begin{equation}}
\def\ee{\end{equation}}
\def\bea{\begin{eqnarray}}
\def\eea{\end{eqnarray}}

\def\vecr{{\bf r}}
\def\lba{\left(}
\def\rba{\right)}
\def\lbc{\left[}
\def\rbc{\right]}

\newcommand{\vp}{{\bf p}}  
\newcommand{\vq}{{\bf q}}  
\newcommand{\vk}{{\bf k}}  

\renewcommand{\vr}{{\bf r}}

\begin{document}

\title{
  Symmetry-breaking Fermi surface deformations 
  \\
  from central interactions in two dimensions
}

\author{Jorge Quintanilla}  
\email{j.quintanilla@rl.ac.uk} 
\affiliation{ISIS Spallation Facility, STFC Rutherford Appleton Laboratory,
Harwell Science and Innovation Campus, Didcot OX11 0QX, United Kingdom}

\author{Masudul Haque} 
\affiliation{Max Planck Institute for Physics of Complex
Systems, N\"othnitzer Str.~38, Dresden, Germany}

\author{A. J. Schofield} 
\affiliation{School of Physics and Astronomy, University of Birmingham,
Birmingham B15 2TT, United Kingdom}

\date{\today}


\begin{abstract}
We present a mean field theory of the Pomeranchuk instability in two
dimensions, starting from a generic central interaction potential described in
terms of a few microscopic parameters. For a significant range of parameters,
the instability is found to be pre-empted by a first-order quantum phase
transition. We provide the ground state phase diagram in terms of our generic
parameters.
\end{abstract}


\maketitle

%
\section{Introduction}
%

A central theme in the study of strongly correlated electron systems is the
appearance of novel types of ordering, and phase transitions leading to such
unconventionally ordered states.  Phases with liquid crystalline symmetry have
emerged as an intriguing theme. They have been studied and proposed, for
example, in the context of quantum Hall systems, 
\cite{FradkinKivelson_QHE-LiqCrysPhases_PRB99, MacDonaldFisher_QHE-smectic,
LapilliWexler_QHE-LiqCrys, WexlerDorsey_QHE-LiqCrys_PRB01,
RadzihovskyDorsey_QHE-nematics_PRL02,
FradkinKivelson_finite-T-nematic_B-field_PRL2000,2002-Ciftja-Wexler,
2007-Doan-Manousakis}
$s$-wave pairing of polarized fermions, 
\cite{KunYang_LiqCrys-AsymmetricPairing_cm05}
and in Hubbard-like models. 
\cite{HalbothMetzner_PRL00, KivelsonFradkinEmery_Nature98,
YamaseOganesyanMetzner_PRB05, DellAnnaMetzner_PRB06,
MetznerRoheAndergassen_PRL03,
KivelsonFradkinGeballe_nematic-in-2DEmery_PRB04}
One prominent mechanism for such phases is via \emph{Pomeranchuk
instabilities}, which are distortion instabilities of the Fermi surface.
\cite{1958-Pomeranchuk} A Pomeranchuk instability occurs in the
angular-momentum channel $l$ when the corresponding Fermi liquid parameter
$F_l$ is sufficiently negative. \cite{1958-Pomeranchuk} The phase resulting
from an $l=2$ type instability is a \emph{nematic}, because the orientaion
symmetry of the continuum or the orientation symmetry of the lattice point
group is broken, modulo an inversion symmetry, while translational symmetry
remains unbroken.  Pomeranchuk instabilities have received significant
attention recently, both in the continuum \cite{2001-OKF,
2006-Quintanilla-Schofield, KunYang_Pomeranchuk-z_cm05,
CastroNeto_z-from-bosonizn_cond-mat2005, Fradkin_z-from-bosonizn_PRB2006,
Kee_continuum_sound_PRB2003,KimKee_continuum-pairing_JPCM2004} and lattice
\cite{YamaseOganesyanMetzner_PRB05, KeeKim_itinerant-metamagnetism_PRB05,
KeeKimChung_nematic-signatures_PRB03, 2004-Khavkine-Chung-Oganesyan-Kee,
DohFriedmanKee_PRB06} contexts.

In this Article we focus on continuum systems, where the Pomeranchuk
instability breaks a continuous symmetry. This is particularly relevant to ultracold fermionic gases and low density 2D electron systems. The Hamiltonian most prominently
studied for this case has been of the quadrupole-quadrupole type explicitly
designed in Ref.~\onlinecite{2001-OKF} to produce an $l=2$
instability. \cite{Kee_continuum_sound_PRB2003,KimKee_continuum-pairing_JPCM2004,
Fradkin_z-from-bosonizn_PRB2006} It is therefore important to study
Pomeranchuk instabilities arising from more generic Hamiltonians.  
In a previous paper, two of the present authors studied shape
deformations of the Fermi surface of a three-dimensional system, arising from
\emph{central} interactions. \cite{2006-Quintanilla-Schofield}  In particular,
Ref.\ \onlinecite{2006-Quintanilla-Schofield} finds that a central
interaction, if it has a sharp feature at a finite length scale $r_0 \gtrsim
k_F^{-1}$, can cause deformations of the Fermi surface. Non-monotonic, "delta-shell" and monotonic, "hard-core" repulsive potentials
were analysed, and they were all found to lead to the effect. A screened
Coulomb interaction, on the other hand, was found not to lead to a Pomeranchuk
instability.

\begin{figure}
\includegraphics[width=1.0\columnwidth]{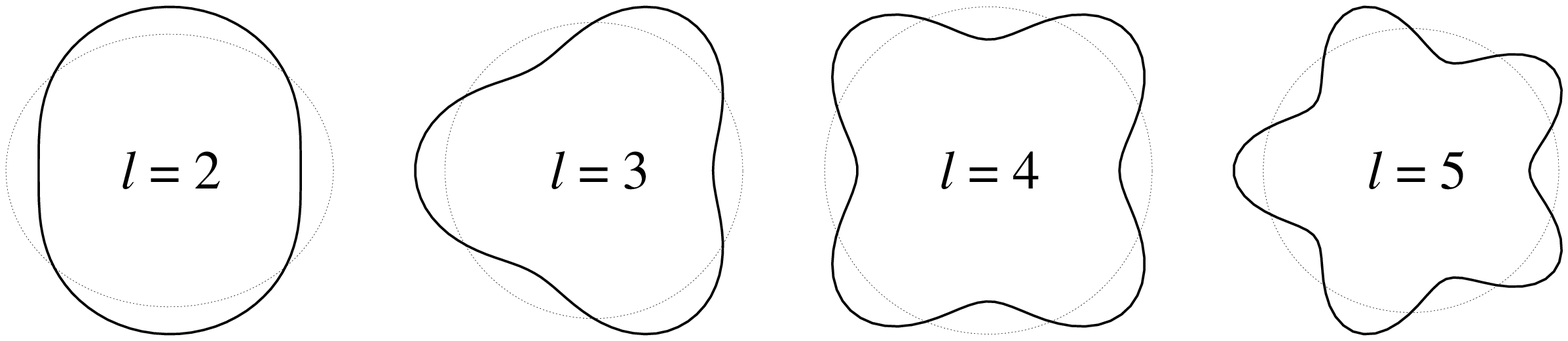}
\caption{Shape of Fermi surface of 2D continuum system, before (thinner line)
  and after (thicker line) a Pomeranchuk instability.  Several possible
  symmetries ($l=2,3,4,5$) are shown for the Pomeranchuk order.
\label{Fig.1}
}
\end{figure}

In this paper we provide a mean-field treatment of two dimensions, where much
of the current interest lies. In the interest of providing generic results, we
will parametrize our central interaction $V(r)$ by a small number of
parameters, namely the values of angular-momentum components of the potential
and its momentum-space derivatives at the Fermi momentum.  These turn out to
be the essential parameters for the description of Fermi surface shape distortion
transtions.  A phase diagram in terms of these parameters provides much more
general information than the consideration of particular forms of
$V(r)$. \cite{2006-Quintanilla-Schofield} Our framework has the added
advantage that a single description treats not only the transition in the
$l=2$ channel, but in \emph{every} angular momentum channel $l>1$.  In other
words, we in fact present a mean field theory not only for distortions leading
to \emph{nematic} symmetry, but also to the others shown in Fig.~\ref{Fig.1}.

One important result of our analysis is that we find the shape-deformation
transitions to be of first order for significant regions of parameter space.
This implies that in many realistic cases where one might get shape
deformation instabilities of the Fermi surface, the transition is discontinuous
and does not involve quantum critical behavior.

We introduce the model in Sec.\ \ref{sec_model} and the mean field treatment
in Sec.\ \ref{sec_MFT}.  The details of the theory are worked out in the next
three sections. Sec.\ \ref{sec_transition-order} describes the
parameter regimes where we have first-order transitions.  In Sec.\
\ref{sec_OKF-comparison} we provide a comparison with the Hamiltonian of
Ref.~\onlinecite{2001-OKF}, which is the dominant Hamiltonian used
in the recent literature for the continuum $l=2$ Pomeranchuk transition.
Finally, in Sec.\ \ref{sec:lan_par} we use our mean field theory to compute the
Landau Fermi liquid parameters in terms of the microscopic interaction
potential. In the final section we lay out our conclusions.

%
\section{Model}  \label{sec_model}
%

We consider the following continuum Hamiltonian, describing interacting
 fermions in two dimensions:
\begin{multline}
H =
  \int d \vecr \psi^{\dagger}_{\vecr} 
  \lbc 
    \epsilon_{\nabla_{\vecr}/i} - \mu
  \rbc 
  \psi_{\vecr}  \\
  +~ \frac{1}{2} \int d\vecr \int d\vecr'
  V(|\vecr-\vecr'|) \psi^{\dag}_{\vecr}\psi^{\dag}_{\vecr'}\psi_{\vecr'}\psi_{\vecr}
  \label{CP1}
\end{multline}
Here $\epsilon_{\hat{\vp}/\hbar}=\hat{\bf p}^2/2m$ gives the `bare' dispersion
relation in terms of the momentum operator $\hat{\vp} \equiv
\lba\hbar/i\rba\nabla_{\vecr}$ (free fermions of mass m). $V(r)$ is a central
interaction potential. For simplicity we consider spinless fermions. (In any
case, when the interaction is spin-independent the Pomeranchuk instability
will be degenerate in the spin channels.  \cite{2006-Quintanilla-Schofield} On the other hand for spin-dependent interactions there can be emergent spin-orbit coupling \cite{2004-Wu-Zhang} and non-trivial spin configurations in momentum space. \cite{2007-Wu-Sun-Fradkin-Zhang})
In reciprocal space, Hamiltonian \eqref{CP1} takes the form \pn{2007-12-26, p32}
\pn{2007-08-20, p1}
\begin{multline}
  H =
  \sum_{\vk} 
  \left(
    \epsilon_{\vk}
    -\mu
  \right)
  \psi_{\vk}^{\dagger}\psi_{\vk}  \\
  +\frac{1}{2\Omega} \sum_{\vk,\vk',\vq}
   V(\vk-\vk')
  \psi_{\vq/2-\vk'}^{\dagger} 
  \psi_{\vq/2+\vk'}^{\dagger} 
  \psi_{\vq/2+\vk}
  \psi_{\vq/2-\vk},
  \label{CP2}
\end{multline}
%
%
where $\psi_{\vk}^{\dagger}=\int d^2 \vr ~ \Omega^{-1/2}e^{i\vk.\vr}
\psi_{\vr}^{\dagger}$ and $V({\bf k})\equiv \int d^2{\bf r} e^{-i{\bf k}\cdot{\bf
r}}V(r)$.  The area of the sample is $\Omega$.

There is a large body of literature on behaviors that can emerge from
Hamiltonians of the type given by Eq.~(\ref{CP2}). For the attractive case
see, for example, the discussions of unconventional superconductivity,
Bose-Einstein condensation of ``preformed pairs'' and ``fermion condensation''
in Refs.~\onlinecite{1997-Stintzing-Zwerger,
2003-Quintanilla-Gyorffy-Annett-Wallington, 1985-Nozieres-SchmittRink,
1992-Nozieres}. Here we are concerned with the emergence of anisotropy of the
Fermi surface due to \emph{repulsive} interactions. As in the case of
anisotropic pairing in superconductors, the key is the dependence of the
interaction strength $V(\vq)$ on momentum transfer, $\hbar\vq$. In the case
of a contact potential, $V(\vq)=\mbox{constant}$, the only possibilities are
isotropic: a conventional Bardeen-Cooper-Schrieffer (BCS) instability for attractive interactions or a
quantum gas-liquid transition for repulsion. 

With more general momentum dependence, Hamiltonian \eqref{CP2} can lead to an
anisotropic state through a Pomeranchuk instability.  The relevant order
parameter is
\[
\langle \hat{b}^{\dagger}(\phi) \rangle
\equiv
\sum_{\vk} \cos \left( l\theta_{\vk} \right) \langle
\psi_{\vk}^{\dagger}\psi_{\vk} \rangle,
\]
where $l=2,3,4,\ldots$, leading to the Fermi surface shape deformations
represented in Fig.~\ref{Fig.1}.  The $l=1$ instability is forbidden in a
Galilean-invariant system.
\cite{2006-Quintanilla-Schofield,2007-Wolfle-Rosch}

%

Pomeranchuk order can be described as particle-hole pairing with
center-of-mass momentum $\vq = 0$ but finite internal angular momentum quantum
number, $l > 0$. In this sense, it is the analog of ``exotic''
superconductivity, where $l>0$ pairing occurs in the particle-particle
channel. Indeed that analogue has been shown recently by one of us to
extend quantitatively to the disorder dependence of a Pomeranchuk transition
temperature.~\cite{2007-Ho-Schofield}

It is useful to introduce the partial wave decomposition of the interaction
potential: \pn{2007-07-13}
\begin{equation}
  V({\bf k}-{\bf k}')=\sum_{l=0}^{\infty} V_l(k,k') 
  \cos\left[l\left(\theta_{\bf k}-\theta_{{\bf k}'}\right)\right]
  \label{pw1}
\end{equation}
where $k=|{\bf k}|$, etc. The  amplitudes $V_l$ are
\begin{multline}
  V_l(k,k') = \left(1+\delta_{l,0}\right)4\pi   \times   \\
\int_0^{\infty} dr r \; J_{l}(kr) J_{l}(k'r) V(r).
  \label{pw2}
\end{multline}
(Note that $V(k,q)=V(q,k)$).
We provide below a mean-field theory that is applicable to any Hamiltonian of
the form \eqref{CP2}.  The theory is quite independent of the details of the
interaction potential $V(|{\vr}-{\vr}'|)$, but depends only on a few
parameters derived from it.  In particular, we will use: (i) the values of the
amplitudes $V_l$ on the Fermi surface (ii) their derivatives on the Fermi
surface.
These are the parameters $V^{(n)}$ introduced in Eq.~(\ref{pre7}).

%
\section{Mean Field Theory}  \label{sec_MFT}
%

%
Our mean-field theory is based on the following {\it ansatz} for the ground
state:
\begin{equation}
   \ket{\Lambda_{\vk}} = 
   \prod_{\vk} 
   \left[
      \Theta\left(
      	 \varepsilon_\vk
      \right)
      +\Theta\left(
      	  -\varepsilon_\vk
      \right)
      \psi^\dag(\vk) 
   \right]
   \ket{0}.
   \label{trial_gs}
\end{equation}
Here $\ket{0}$ is the vacuum and $\psi^{\dagger}(\vk)$ creates an electron in
a plane wave state.  The {\it ansatz} wavefunction is a Slater determinant of
plane waves. $\varepsilon_{\vk}$ is an \emph{arbitrary} dispersion relation.
Its form dictates which plane wave states are occupied:
  \begin{equation}
    \varepsilon_{\vk}
    \equiv
    \epsilon_{|\vk|}-\mu-\Lambda_\vk \leq 0,
    \label{b3}
  \end{equation}
where we have introduced a `mean field' $\Lambda_\vk$ which is the difference
between the `bare' dispersion relation $\epsilon_{k}-\mu$ and the renormalized
one.

The mean field $\Lambda_\vk$ is our variational parameter.  This field
renormalizes the bare electronic dispersion relation, and therefore the Fermi
surface, which is defined as $\varepsilon_{\vk}=0$.  Minimizing the ground
state energy,
\begin{multline}
  E=\bra{\Lambda_{\vk}}H\ket{\Lambda_{\vk}}
  = \sum_{\vk} \left( \epsilon_{\vk} - \mu \right) N_{\vk}
   \\ + \frac{1}{2\Omega} \sum_{\vk,\vk'}
    N_{\vk} N_{\vk'} 
    \left[
      V(0) - V(\vk-\vk')
    \right],
    \label{E-1}
\end{multline}
with respect to the
functional form of $\Lambda_{\vk}$ yields a self-consistency equation that
determines $\Lambda_\vk$. Proceeding as in
Ref.~\onlinecite{2006-Quintanilla-Schofield}, we find \pn{2007-08-20, p2}
\begin{equation}
  \Lambda_\vq = \frac{1}{\Omega}
    \sum_{\vk} \left[
      V\left(\vk-\vq\right)-\bar{V}
    \right] N_{\vk}
  ,
  \label{self_cons}
\end{equation}
where $\bar{V} \equiv \int d^2\vecr V(r)$ and
$N_{\vk}\equiv\xpct{\psi_{\vk}^{\dagger}\psi_{\vk}}=\Theta\left(-\varepsilon_\vk\right)$. Note that $\Lambda_\vq$ coincides with the one-loop, Hartree-Fock approximation to the Fermionic self-energy $\Sigma(\vq,\omega)$ (the frequency-dependence dropping out at this level, for static interactions). When
the above equation has more than one solution, the one that minimises
\begin{equation}
  E = \sum_\vk 
    N_\vk
    \left(
      \varepsilon_{\vk}+\frac{1}{2} \Lambda_\vk 
    \right) \; 
  \label{E_stat}
\end{equation}
must be chosen. The above expression results from substituting Eq.~(\ref{self_cons}) into Eq.~(\ref{E-1}). The second term inside the brackets may be interpreted as a
double-counting correction to the naive mean-field theory which emerges from
the variational calculation.


\section{\label{sec:small_def}Small Fermi surface deformations}

Let us split the mean field $\Lambda_{\vk}$ into
two parts: a rotation-symmetric part and a symmetry-breaking part.  The latter
will have a number of components, corresponding to different values of the
angular momentum quantum number of the electron-hole pair. Nevertheless near
an instability of the isotropic state, or a sufficiently weak first-order transition out of it, we can assume, save accidental
degeneracies, that only one of these components is finite. We thus write
\begin{equation}
  \Lambda_{\vq} = 
  \Lambda_0(q)+\Lambda_l(q)\cos(l\theta_{\vq}),
  \label{one_l}
\end{equation}
or, equivalently,
\be
\varepsilon_{\vq}
=
\varepsilon_{0}(q)-\Lambda_l(q)\cos(l\theta_{\vq}),
\label{one_l_2}
\ee
where $\varepsilon_{0}(q) \equiv \hbar^2q^2/2m-\mu-\Lambda_0(q)$ is
the renormalized dispersion relation before the instability sets in and
$l=2,3,4,\ldots$ determines the symmetry of the instability (see
Fig.~\ref{Fig.1}).  We have chosen a particular orientation of the
deformation of the Fermi surface, without loss of generality.

In the symmetric
phase (zero deformation), the Fermi momentum $\hbar{k_F^0}$ 
is
defined by
\[
  \varepsilon_0(k_F^0)=0  
.
\]
In the symmetry-broken phase, 
this quantity depends
on the direction of
${\vk}$, given by the angle $\theta$:
\begin{eqnarray}
  k_F^0
  & ~\to~ &
 k_F(\theta) ~=~  k_F^0 + \delta k_F ( \theta ) 
  .
\end{eqnarray}
The offset of the Fermi vector is given by
\be
	\varepsilon_0\left(k_F^0+\delta k_F\right) 
     - \Lambda_l 
     \left( 
     	k_F^0+\delta k_F
     \right) 
     \cos \left( l\theta \right) 
     = 0
     \label{11}
\ee
All results presented in the remaining of this paper have been obtained by solving this equation for small deformations of the Fermi surface, i.e. under the assumption that 
\begin{equation}
  |\delta k_F(\theta)|  \ll  k_F^0
  \label{2a}
\end{equation}
%
in all directions $\theta$. In particular we will assume $\delta k_F$ to be small enough that the symmetric part of the dispersion relation can be linearised:
\be
	\varepsilon_0\left(k_F^0+\delta k_F^0\right)
     \approx
     \hbar v_F^0 \delta k_F.
     \label{X}
\ee
This is quite distinct from the work of other authors, where non-linear terms in the dispersion relation were invoked to stabilize a quantum critical point.  
\cite{2001-OKF,2003-Barci-Oxman,2006-Lawler-Barci-Fernandez-Fradkin-Oxman,2008-Barci-Trobo-Fernandez-Oxman} That is discussed in detail in Sec.~\ref{sec_OKF-comparison}. Similarly we will assume that, within $\delta k_F^0$ of the Fermi vector, the $|{\bf k}|$-dependent amplitude of the deformation potential can be approximated by a constant:
\be
	\Lambda_l\left(k_F^0+\delta k_F\right) 
     \approx
     \Lambda_l\left(k_F^0\right)
     \equiv \Lambda
     \label{Y}
\ee
Note that we are not writing explicitely the dependence of $\Lambda$ on the angular momentum quantum number, $l$.

The approximations (\ref{X},\ref{Y}) are valid when the following conditions hold for $n=1,2,3,\ldots$:
\be
  \frac{1}{(n+1)!} \varepsilon_0^{(n+1)}\delta k_F^{n} 
  ,
  \frac{1}{n!} \Lambda^{(n)}\delta k_F^{n-1} 
  \ll
  \hbar v_F^0,
  \label{last_approx_1}
\ee
where ${\Lambda}^{(n)}\equiv\partial^n \Lambda_l(k)/\partial k^n|_{k=k_F^0}$ and  ${\varepsilon}_0^{(n)}\equiv\partial^n \varepsilon_0(k)/\partial k^n|_{k=k_F^0}$. 
Under these conditions, Eq.~(\ref{11}) yields the deformation of the Fermi surface as
\be
	\delta k_F(\theta) 
	=
     \frac{\Lambda}{\hbar v_F^0}\cos(l\theta).
	\label{3}
\ee
Note that Eq. (\ref{3}) allows us to re-write the small-deformation condition (\ref{2a}) in the form 
\be
	\Lambda \ll \hbar \k_F^0 v_F^0.
  \label{cond}
\ee

Using the decompositions \eqref{pw1} and \eqref{one_l} in the
self-consistency equation (\ref{self_cons}), we can express the self-consistency equation as 
\begin{multline}
  \frac{1}{4\pi^2} 
        \int_0^{2\pi} d\theta_{\bf k} \cos(l\theta_{\vk})
        \int_{k_F^0}^{k_F^0+\delta k_F(\theta)} dk \; k \;
 V_l(k, k_F^0)
\\ =
 \Lambda.
 \label{4b1}
\end{multline}
In general, this is a self-consistency equation determining the values of $\Lambda$ that minimize (and maximize) the energy. The self-concistency comes in through the dependence of $\delta k_F^0 (\theta)$ on $\Lambda$, Eq.~(\ref{3}). 
Now, the integral equation (\ref{4b1}) can be reduced to a polynomial equation in $\Lambda$ by invoking condition (\ref{11}) again to keep a finite number of terms of the expansion of the interaction potential around the Fermi surface:
\begin{multline}
V_l(k, k_F^0) ~\approx~ V + V'\left( k -k_F^0\right) \\
+ \frac{1}{2}V''\left(k -k_F^0\right)^2  \\
+ \frac{1}{3!}V'''\left( k-k_F^0\right)^3
+ \ldots
\label{pre7}
\end{multline}
Here $V \equiv V_l(k_F^0,k_F^0)$ gives the strength of the coupling on the
Fermi surface and $V' \equiv \partial V_l(k,k_F^0)/\partial k|_{k=k_F^0} =
\partial V_l(k_F^0, k)/\partial k|_{k=k_F^0}$ is its slope. Higher derivatives
give the curvature, etc.  
Substituting Eq.~(\ref{pre7}) into Eq.~(\ref{4b1}) we obtain the following
equation:
\begin{multline}
  \Lambda
  =
  \frac{1}{4\pi^2} 
  \left\lbrace
  Vk_F^0
  \int_0^{2\pi} d\theta \cos(l\theta) \delta k_F(\theta) 
\right.
	\\
	+ 
	\frac{V+V'k_F^0}{2}
	\int_0^{2\pi} d\theta \cos(l\theta) \delta k_F(\theta)^2
  \\
  \left.
	  + 
	  \frac{V'}{3}
	  \int_0^{2\pi} d\theta \cos(l\theta) \delta k_F(\theta)^3
	  + \ldots
  \right\rbrace
  \label{L_ord}
\end{multline}
Substituting the value of $\delta k_F(\theta)$ given by Eq.~(\ref{3}) and carrying out the integration with respect to $\theta$ we obtain, to a given order in $\delta k_F/k_F^0 = \Lambda/\hbar k_F^0 v_F^0$, a polynomial equation in $\Lambda$.

\section{\label{sec:inst_eq}Instability equation}

The instability equation is found by solving the self-consistency equation (\ref{4b1}) at lowest order in our small parameter expansion, Eq.~(\ref{pre7}).
Using $V_l(k,k_F^0)\approx V_l(k_F^0,k_F^0)$, the critical $V_l$
required for the instability is found to be \pn{2007-08-20,p.4}
\begin{equation}
V_l
(k_F^0,k_F^0) = \frac{4\pi \hbar v_F^0}{k_F^0} \equiv V_{\rm crit}.
\label{6}
\end{equation}
%
%
This is our instability equation.  Note that $V_l(k_F^0,k_F^0)$ is the only
parameter of the interaction potential entering the instability equation
(although the $l=0$ amplitude is also important, as it renormalises the
Fermi velocity $v_F^0$). 
For $V_l(k_F^0,k_F^0) < V_{\rm crit}$, the symmetric Fermi surface is a
(local) energy minimum. For $V_l(k_F^0,k_F^0) > V_{\rm crit}$, it is a
maximum.

Writing $V_l(k_F^0,k_F^0)$ and $v_F^0$ explicitly in the above equation, we find it in the form
\pn{2007-08-20,p.9}
\begin{equation}
  \frac{\hbar^2}{m}
  =
  \int_0^{\infty}
  drrV(r)\left[J_l(k_F^0r)^2-J_1(k_F^0r)^2\right]
  \label{inst_eq}
\end{equation}
for $l=1,2,3,\ldots$.  
%
%
Note that the above equation lacks any solutions with $l=1$, as
was found in $D=3$, \cite{2006-Quintanilla-Schofield} due to the Galilean
invariance of the system. \cite{2007-Wolfle-Rosch}

As discussed in Ref.~\onlinecite{2006-Quintanilla-Schofield} for the
three-dimensional case, Eq.~\eqref{6} is a microscopic version of the
Pomeranchuk instability condition, \cite{1958-Pomeranchuk} and Eq.~\eqref{6}
reduces to the Pomeranchuk condition if we use our mean field theory to
compute the Landau Fermi liquid parameters in terms of the microscopic
interaction $V(|\vk-\vk'|)$ (see Sec.~\ref{sec:lan_par}).

%
\section{\label{sec:ordered-state}Ordered state}  
%

\begin{figure}
\includegraphics{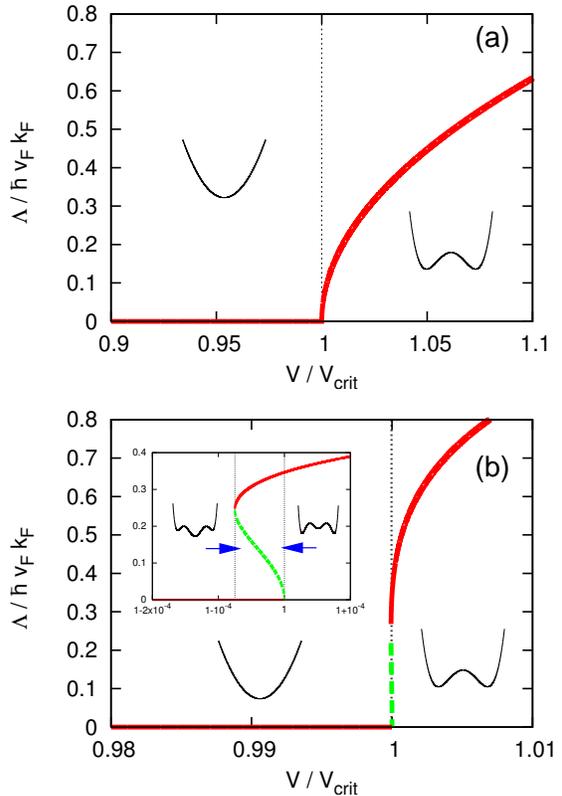}
\caption{
\label{Fig.2} Values of the amplitude on the Fermi surface of the deformation
potential, $\Lambda$, that minimise the free energy (solid lines), plotted as
functions of the coupling strength in units of its ``critical'' value,
$V/V_{\rm crit}$. (a) $V'k_F/V_{\rm crit} = -1$ and $V^{(n)}k_F^n/V_{\rm
  crit} \approx 0$ for all $n=2,3,\ldots$ [Eq.~(\ref{10a2})]
(b) $V'k_F/V_{\rm crit} = 1$, $V^{2}k_F^2/V_{\rm crit} = -2+1/50$,
$V^{3}k_F^3/V_{\rm crit} = -1$ and $V^{(n)}k_F^n/V_{\rm crit} \approx 0$
for $n \geq 4$ [Eq.~(\ref{11a2_dless})]. The dashed line in panel (b)
indicates an additional stationary point, but it is a maximum, not a
minimum. The region where the transition takes place has been blown out in the inset. The dotted lines indicate the critical coupling of Eq.~(\ref{6})
[panels (a) and (b)] and the lower bound in Eq.~(\ref{Vjump}) [panel (b)]. The
structure of the free energy in the different parameter regions has been
sketched for illustration.  }
\end{figure}

We now discuss the evolution of the amplitude of the deformation, $\Lambda$,
in the ordered ground state realised when $V_m(k_F,k_F) > V_{\rm crit}$. To describe the ordered state, we will need to go beyond the lowest order in $\delta k_F / k_F^0$ in  Eq.~(\ref{pre7}).
To derive the instability equation \eqref{6}, we
only used the first parameter in the expansion of the
$l^{\rm th}$ component of the interaction potential, Eq.~(\ref{pre7}): $V_l(k,k_F^0)\approx V$.  Let us now also keep the first derivative:
\begin{equation}
V_l(k,k_F^0)\approx V + V'\left(k-k_F^0\right).
\label{7}
\end{equation}
When substituted in the self-consistency equation in the form of
Eq.~\eqref{4b1}, we get, after integrating with respect to $k$, \pn{2007-10-12}
Eq.~(\ref{L_ord}).
Substituting Eq.~(\ref{3}) and carrying out the integrals we get
\pn{2007-11-26b}
\begin{equation}
  \Lambda
  =
  V\frac{k_F^0}{4\pi \hbar v_F^0} 
  \Lambda
  + \frac{V'}{16\pi \hbar^3 {v_F^0}^3} 
  \Lambda^3
  .
  \label{L_ord_2}
\end{equation}
This equation
admits two solutions: the trivial one, 
\(
\Lambda = 0,
\label{10a1}
\) 
minimises the ground state energy when $V < V_{\rm crit}$.  On the other
hand when $V > V_{\rm crit}$ the above solution is a maximum. If $V'<0$, the minima are at
\pn{2007-10-15, p.12}
\begin{equation}
  \frac{\Lambda}{\hbar v_F^0 k_F^0}
  =
  \pm
  \left(
    -\frac{4 V_{\rm crit}}{V'k_F^0}
  \right)^{1/2}
  \left(
    \frac{V-V_{\rm crit}}{V_{\rm crit}}
  \right)^{1/2}.
  \label{10a2}
\end{equation}
The amplitude on the Fermi surface of the deformation potential thus grows in
second-order fashion, with critical exponent $= 1/2$, as expected for this
mean-field theory and depicted in Fig.~\ref{Fig.2}~(a). If $V'>0$, on the other
hand, we obtain the unphysical 
result that the Fermi surface deformation decreases as the instability point is approached from below [see Fig.~\ref{Fig.2}~(b)]. This is an indication that linearizing the interaction in
Eq.~(\ref{pre7}) is no longer adequate to describe a transition which
potentially is becoming first-order. To address this we go beyond the
assumption of Eq.~(\ref{7}) in the next section.
%

%
\section{First-order quantum phase transition}    \label{sec_transition-order}

%

Within our linearized theory we have shown that, if $V'>0$, unphysical solutions occur. These solutions
are suggestive of the isotropic state being a local minimum with the
true ground state separated from it by a first order transition where
the Fermi surface jumps to one of lower symmetry.  Linearisation of
$V_l(k,k_F^0)$ around $k=k_F^0$, Eq.~(\ref{7}), is therefore not
realistic for change that is no longer infinitesimal near the
transition.
We therefore 
carry out the expansion (\ref{pre7}) to the third, rather than first,
order. 
It is useful to introduce the following, dimensionless parameters describing the interaction potential and the resulting deformation of the band structure:
\begin{eqnarray}
  v^{(n)} 
  & \equiv &
  \frac{V^{(n)}{k_F^0}^n}{V_{\rm crit}}
  \label{vns} 
  \\ 
  \lambda & \equiv & \frac{\Lambda}{\hbar v_F^0 k_F^0}.
\end{eqnarray}
In terms of these, the self-consistency equation takes the form \pn{2007-10-15, p.12}
\begin{equation}
  (v-1)  \lambda 
  +
  \frac{1}{4}
  \left(
    v'+\frac{v''}{2}
  \right)
  \lambda^3 
  +
  \frac{v'''}{48}
      \lambda^5  
  =
  0. 
  \label{11a2_dless}
\end{equation}
Taking $v'' \approx v''' \approx 0$ and solving for $\lambda$ recovers Eq.~(\ref{10a2}) as
\(
  \lambda
  =
  \pm 2\left(-1/v'\right)^{1/2}\left(v-1\right)^{1/2}. 
\) 
However, Eq.~(\ref{11a2_dless}) has valid solutions for $v' > 0$, too. All the
solutions are straightforward to obtain analytically. Two examples are plotted in
Fig.~\ref{Fig.2}.

\begin{figure}
\includegraphics{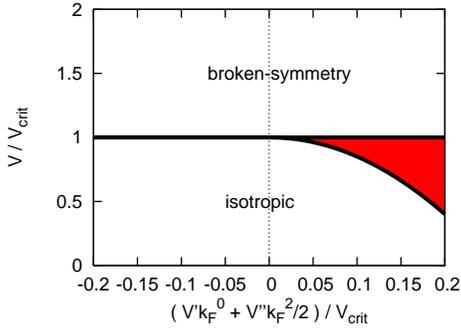}
\caption{\label{Fig.3}Phase diagram for small, symmetry-breaking deformations
of the Fermi surface with a given value of the angular momentum quantum number
$l$. In the symmetric state, the Fermi surface is circular. In the
broken-symmetry state, it has one of the configurations of
Fig.~\ref{Fig.1}. In the shaded region one of the two states is stable, and
the other meta-stable, so the transition takes place through a first order
jump. The parameter $V'''k_F^3/V_{\rm crit} = -1/20$, which controls the width
of this region [Eq.~(\ref{Vjump})].}
\end{figure}

We note that there are two types of behaviour. For 
\begin{equation} v'+v''/2<0 \label{2nd_order} \end{equation} 
(which reduces to $v'<0$ in the limit of small $v''$) there is a second-order
transition as we found above.  However, when the above condition is not met,
there is a range of values of $V$ for which the free energy has a triple-well
structure.  The instability is then pre-empted by a weakly first-order quantum
phase transition: a small ``jump'' in the shape of the Fermi surface. The
value of the coupling strength at which this happens, $V_{\rm jump} < V_{\rm
crit}$, is bounded by \pn{2007-10-15, p.23}
\begin{equation}
  1 + \frac{3\left(v'+v''/2\right)^2}{4v'''}
  \leq
  \frac{V_{\rm jump}}{V_{\rm crit}}
  <
  1.
  \label{Vjump}
\end{equation}
The corresponding phase diagram is shown in Fig.~\ref{Fig.3}. The exact
location of the jump depends on additional parameters characterising the
fermion-fermion interaction \pn{2007-10-18, p.36} (Specifically, it
depends on the off-diagonal second and third derivatives of $V_l(k,q)$
with respect to $k$ and $q$.)

The above is only valid for $v''' < 0$. If $v'''>0$, higher-order terms
describing the dependence of $V_l(k,q)$ on $k$ and $q$ even
further away from the Fermi surface become important. In that case the
transition is no longer even weakly first-order and the precise behaviour of the model depends on more
specific details. Naturally as the whole solution relies on the assumption
that $\delta k_F(\theta)$ is small the range in which it is reliable is
restricted to values of the parameters for which the jump is small.

It is important to stress that the above arguments rely on the approximation
of Eq.~(\ref{3}). In the present case, however, the approximation made in that
equation requires not only that $|\Lambda'| \ll v_F$, but indeed that Eq.~(\ref{last_approx_1}) holds
for $n=1,2,3$. Equivalently we have neglected the additional
renormalisation of the Fermi velocity and higher-order derivatives of the
dispersion relation resulting from the symmetry-breaking part of the Fermi
surface deformation, compared to the symmetric contribution.

%
\section{\label{sec:lan_par}Landau parameters}
%

Variation of Eq.~(\ref{E-1}) gives the change of the ground state energy associated with an arbitrary re-distribution of the fermions in momentum space, $N_{\vk} \to N_{\vk}+\delta N_{\vk}$:
\begin{multline}
  \delta E = \sum_{\vk} \varepsilon_{\vk} \delta N_{\vk}
  +\frac{1}{2\Omega} \sum_{\vk,\vk'} f(\vk,\vk')
  \delta N_{\vk} \delta N_{\vk'}.
\end{multline}
This coincides with the usual expression from Landau Fermi liquid theory. \footnote{See, for example, A. J. Leggett, Rev. Mod. Phys. {\bf 47}, 331 (1975).} The effective interaction between quasiparticles is given by $f\left(\vk,\vk'\right) = 
\bar{V} - V\left(\vk-\vk'\right)$ in terms of the microscopic parameters. The Landau Fermi liquid parameters can be defined from this function in the usual way:
\begin{multline}
  f(\vk,\vk') = \frac{1}{\rho(0)} \sum_{l=0}^{\infty} F_l 
  \cos\left[l\left(\theta_{\vk}-\theta_{\vk'}\right)\right]
  \\ \mbox{ for }k=k'=k_F^0
\end{multline}
where $\rho(0) = 1/2\pi \hbar v_F^0$ is the density of states at the Fermi energy. Thus 
\be
  F_l=\rho(0)\left(\delta_{l,0}\bar{V}-V_l\right).
  \label{LFLp_formula}
\ee
In terms of these Landau parameters, the Pomeranchuk instability equation (\ref{6}) of section \ref{sec:inst_eq} takes the usual form, $F_l < -2$. The microscopic parameters $V'$, $V''$, etc. introduced in Sec.~\ref{sec:ordered-state} are related to derivatives of $f(\vk,\vk')$ at the Fermi energy and can thus be regarded as generalisations of the Landau parameters.

%
%
\section{Quadrupole-quadrupole interactions}  \label{sec_OKF-comparison}
%

In Ref.~\onlinecite{2001-OKF} Oganesyan, Kivelson and Fradkin (OKF) introduced an effective
Hamiltonian which has been employed widely to study the $l=2$ instability on the
continuum.  \cite{Kee_continuum_sound_PRB2003,KimKee_continuum-pairing_JPCM2004} It features an anisotropic, ``quadrupole-quadrupole'' effective interaction. By contrast our Hamiltonian, Eq.~(\ref{CP1}), features a central interaction potential. In this Section we relate the two. 

The OKF Hamiltonian is
\begin{multline}
  \label{eq_OKF_ham}
  H_{OKF}
  =
  \int d\vr \psi^{\dagger}(\vr) \lbc
  -\frac{\hbar^2\nabla^2}{2m} - \mu
  \rbc \psi(\vr)  \\
  +\frac{1}{4} \int d\vr \int d\vr' {\mathcal F}_2(\vr-\vr')
  \tr[\hat{\bf Q}^{\dag}(\vr)\hat{\bf Q}(\vr')].
\end{multline}
This corresponds to postulating directly an anisotropic density-density
interaction with $l=2$ symmetry which, by analogy with classical liquid
crystals, is assumed to be of quadrupole-quadrupole form.  \cite{C&W} The ``quadrupoles'' used here are the quadrupole moments of the electronic momentum distribution, \cite{2001-OKF}
\begin{equation*}
  \hat{\bf Q}(\vr) \equiv
  -\frac{1}{k_F^2}
  \psi^{\dagger}(\vr)
    \begin{pmatrix}
      \hat{\partial}_x^2 - \hat{\partial}_y^2 
      & 
      2 \hat{\partial}_x \hat{\partial}_y
      \\
      2 \hat{\partial}_x \hat{\partial}_y
      &
      \hat{\partial}_y^2 - \hat{\partial}_x^2       
    \end{pmatrix}
  \psi(\vr)    \; .
\end{equation*}
%

The philosophy behind the effective Hamiltonian in
Eq.~(\ref{eq_OKF_ham}) is that in order to describe the important
fluctuations on approaching the Pomeranchuk instability, in a given
channel $l$, from the isotropic state it is not important to include
fluctuations tending to create either distortions of the Fermi surface
shape with different symmetry or an inhomogeneous state
(backscattering terms). 
To understand the relationship of theories based on this approximation to our analysis of the generic central
Hamiltonian of Eq.~\eqref{CP2}, we will construct a central Hamiltonian whose form is
constrained so that its leading instability has an effective
Hamiltonian of the OKF form.

In order to compare with the generic Hamiltonian \eqref{CP2}, we first move to
reciprocal space:
\begin{equation*}
  \hat{\bf Q}(\vr) \equiv
  \sum_{\vk_1\vk_2} \psi^\dag (\vk_1) Q_{\vk_2} 
  \psi(\vk_2) e^{i(\vk_1-\vk_2).\vr}
  \; .
\end{equation*}
One finds $Q_{\vk}^{11} = - Q_{\vk}^{22} = -(k^2/k_F^2)\cos(2\theta)$, and
$Q_{\vk}^{12} =  Q_{\vk}^{21} = (k^2/k_F^2)\sin(2\theta)$, so that 
\(
\label{Qk3}
\tr [Q_{\vk_1} Q_{\vk_2} ]  = 2 \lba k_1^2 k_2^2/k_{\rm F}^4\rba \cos
2(\th_1-\th_2).\)  
Thus \pn{2007-07-11}
\begin{widetext}
\begin{multline}
  H_{OKF} =
  \sum_{\vk} (\tilde{\epsilon}_{\vk}-\mu) \psi_{\vk}^{\dagger}\psi_{\vk}
  - \frac{1}{2\Omega}\frac{(2\pi)^4}{k_F^4} 
  \sum_{\vk,\vk',\vq} 
  {\mathcal F}_2(\vq)\left(\vk-\frac{\vq}{2}\right)^2\left(\vk'+\frac{\vq}{2}\right)^2
  \\
  \cos \left[ 
    2 \left( 
      \theta_{\vk-\vq/2} - \theta_{\vk'+\vq/2} 
    \right)
  \right]
  \psi_{\vk+\vq/2}^{\dagger} 
  \psi_{\vk'-\vq/2}^{\dagger} 
  \psi_{\vk-\vq/2}
  \psi_{\vk'+\vq/2}
  \label{eq_OKF_ham_kspace}
\end{multline}
%
%
where the Fourier transform of the interaction potential is defined by
${\mathcal F}_2(\vr-\vr')\equiv \Omega^{-1} \sum_{\vq}e^{i\vq.(\vr-\vr')}{\mathcal F}_2(\vq)$ and we have introduced the notation 
\begin{equation}
  \tilde{\epsilon}_{\vk} 
  \equiv 
  \epsilon_{\vk} 
  + \frac{(2\pi)^4}{2\Omega k_F^4} \;
  \sum_{\vq}{\mathcal F}_2\left(\vq\right)
  \left(\vk-\vq\right)^2\vk^2
  \cos
  \left[
    2\left(
      \theta_{\vk-\vq}-\theta_{\vk}
    \right)
  \right].
\end{equation}

Like our Hamiltonian, Eq.~(\ref{eq_OKF_ham_kspace}) features a pairwise
interaction that preserves the total momentum of the pair, $\vq.$ Note,
however, the complicated dependence on $\vq$, $\vk'$ and $\vk$. By contrast in
Eq.~(\ref{CP2}) the interaction depends only on the transferred momentum, $\hbar({\bf k}-{\bf k}')$.  
The dependence on
${\bf q}$, in particular, means that the OKF interaction is not uniform. However the
actual form of ${\mathcal F}_2(\vq)$ is not very important
\cite{2001-OKF} and it is
customary\cite{2001-OKF,2004-Khavkine-Chung-Oganesyan-Kee}
to take ${\mathcal F}_2(\vq)={\mathcal F}_2 \delta_{\vq,0}$. With this assumption
Eq.~(\ref{eq_OKF_ham_kspace}) takes the simpler form
\begin{equation}
  H_{OKF} ~=~
  \sum_{\vk} (\tilde{\epsilon}_{\vk}-\mu) \psi_{\vk}^{\dagger}\psi_{\vk}
  ~-~ \frac{1}{2\Omega}\frac{(2\pi)^4}{k_F^4}  \;
  \sum_{\vk,\vk'} 
  {\mathcal F}_2\vk^2\vk'^2
  \cos \left[ 
    2 \left( 
      \theta_{\vk} - \theta_{\vk'} 
    \right)
  \right]
  \psi_{\vk}^{\dagger} 
  \psi_{\vk'}^{\dagger} 
  \psi_{\vk}
  \psi_{\vk'}
\label{eq_OKF_ham_kspace_qzero}
\end{equation}
\end{widetext}
Let us now compare this Hamiltonian to the form considered here, given by Eq.~(\ref{CP2}).
In our trial ground state, Eq.~(\ref{trial_gs}), the only terms in the triple sum that contribute to the energy have $\vk = \vk'$ or $\vk=-\vk'$. \pn{2007-12-26, pp.33} Omitting all other terms from Eq.~(\ref{CP2}) it takes the simpler form \pn{2008-01-14b}
\begin{multline}
  H =
  \sum_{\vk} 
  \left(
    \epsilon_{\vk}
    -\mu
  \right)
  \psi_{\vk}^{\dagger}\psi_{\vk}  \\
  +\frac{1}{2\Omega} \sum_{\vk,\vk'}
  \left[ V(\vk'-\vk) - V(0) \right] 
  \psi_{\vk}^{\dagger} 
  \psi_{\vk'}^{\dagger} 
  \psi_{\vk}
  \psi_{\vk'}.
  \label{CP2r2}
\end{multline}
Substituting in this expression the partial wave expansion of the interaction potential, Eq.~(\ref{pw1}), we obtain a series of different interaction terms, labelled by $l$, which lead to Fermi surface deformations with different symmetries. Near a second- or weakly first-order Pomeranchuk distortion with $l=2$, only the corresponding term need be considered (see above). Neglecting all the others the Hamiltonian takes the form 
\begin{multline}
  H =
  \sum_{\vk} 
  \left(
    \epsilon_{\vk}
    -\mu
  \right)
  \psi_{\vk}^{\dagger}\psi_{\vk}  \\
  +\frac{1}{2\Omega} \sum_{\vk,\vk'}
  V_2(k,k') \cos \left[ 2 (\theta_{\vk}-\theta_{\vk'}) \right]
  \psi_{\vk}^{\dagger} 
  \psi_{\vk'}^{\dagger} 
  \psi_{\vk}
  \psi_{\vk'},
  \label{CP3}
\end{multline}
whose interaction part coincides with that in Eq.~(\ref{eq_OKF_ham_kspace_qzero}) if we take
%
%
%
%
%
\begin{equation}
  V_{l}(k,k') = -\frac{(2\pi)^4 {\mathcal F}_2}{2k_F^4}
  {k}^2{k'}^2. 
  \label{F2_k}
\end{equation}
Thus although the OKF Hamiltonian does not correspond to a central interaction potential, within our theory it would give the same results as a hypothetical central interaction, whose $l=2$ component happens to be given by Eq.~(\ref{F2_k}). Note that inserting Eq.~(\ref{LFLp_formula}) into Eq.~(\ref{F2_k}) we obtain ${\mathcal F}_2 \propto F_2$, as expected.\cite{2001-OKF} 

From this result we note that, since ${\mathcal F}_2$ is negative at the instability, it follows from Eq.~(\ref{F2_k}) that
\begin{equation}
  V' = -\frac{(2\pi)^4 {\mathcal F}_2}{k_F} > 0,
\end{equation}
(and that $V'' > 0$ too). This implies that within our mean field theory, based on a linearised dispersion relation, Eq.~(\ref{X}), the central potential model which captures the OKF Hamiltonian is in the parameter regime where the
Pomeranchuk instability is actually first order. Indeed to stabilise a quantum critical point for the OKF Hamiltonian it is \emph{essential} to include non-linear terms of the symmetric dispersion relation, $\varepsilon_0(k)$, as noted in Refs.~\onlinecite{2001-OKF,2003-Barci-Oxman,2006-Lawler-Barci-Fernandez-Fradkin-Oxman,2008-Barci-Trobo-Fernandez-Oxman}. Conversely, our complementary approach shows the generic conditions under which a quantum critical point can be stabilised without invoking such non-linearities of the dispersion relation. This is achieved instead by properly taking into account ultra-violet cutoffs implicit in any given central interaction potential, $V(r)$. 

Our results are consistent with other work considering specific microscopic realizations of Fermi surface instabilities which also found wide regions where the transition is first
order~\cite{2004-Khavkine-Chung-Oganesyan-Kee,YamaseOganesyanMetzner_PRB05}. We stress that the main lesson one should extract from this is that the order of the quantum phase transition is a very delicate issue, depending on fine details of the effective interaction and the band structure. Indeed the higher-order terms of the dispersion relation alluded to above modify the coefficients of $\lambda^3$ and $\lambda^5$ in Eq.~(\ref{11a2_dless}) [for a detailed analysis, see Ref.~\onlinecite{2003-Barci-Oxman}.] For example, a large enough $\delta k_F^3$ term in the expansion of $\varepsilon_0(k_F^0+\delta k_F)$ can change the sign of the coefficient of $\lambda^3$, and hence the order of the transition. Thus a full analysis going beyond present calculations would have to treat non-linearities in the interaction and the dispersion relation on an equal footing.\cite{barci} Moreover higher-order effects beyond the scope of mean field theories may well upset this balance one way or the other. Such non-mean field effects will certainly become important whenever the condition (\ref{cond}) does not hold ---e.g. if the phase transition is of first order, but not weakly so.

%
\section{Conclusion} 
%

We have provided a mean-field theory for continuum Pomeranchuk transitions in
two dimensions.  The theory is expressed in terms of a few pertinent
parameters ($V$, $V'$, $V''$, ...) for each angular momentum channel.  This
makes the theory quite general, and applicable for a wide class of central
interactions in which the symmetry breaking is not put in explicitly by hand.

Our main results are Eqs.~(\ref{2nd_order}) and (\ref{Vjump}), which determine the phase diagram in Fig.~\ref{Fig.3}. They apply to any central interaction potential in a two-dimensional continuum, described in terms of a few dimensionless parameters, defined in Eq.~(\ref{vns}). Depending on the form of the interaction, our theory may lead either to a first- or second-order quantum phase transition. Thus our approach is complementary to other work \cite{2001-OKF,2003-Barci-Oxman,2006-Lawler-Barci-Fernandez-Fradkin-Oxman,2008-Barci-Trobo-Fernandez-Oxman}  where a quantum critical point was stabilised by non-linear terms in the dispersion relation.

A continuum theory is useful for several reasons.  One is the direct relevance
to several experimental systems where a Pomeranchuk transition might be
realized, for example: 2D electron layers at semiconductor heterojunctions or
on liquid helium; layered helium systems \cite{1995-Godfrin-Lauter} (where a new phase intervening between the Fermi liquid and Mott insulator states has been observed\cite{2007-Neumann-Nyeki-Cowan-Saunders}); or in a cold-atom setting with trapped fermionic atoms with
dipolar repulsion. These are situations where underlying lattice structures
are not expected to play a role.  Theoretically, continuum Pomeranchuk
transitions are fascinating because the resulting broken symmetries are
remarkable.  While some effects of the nematic symmetry (broken $O$(2)/$Z_2$)
have already been explored, \cite{2001-OKF} we believe there are further
implications, for example, effects of nematic half-vortex excitations in such
a medium.\cite{1978-Stein}  Our framework also provides a description of transitions in
higher-$l$ channels (leading to $O$(2)/$Z_l$ symmetry broken states), which
presumably leads to a broader class of interesting excitations and properties.
Some of these issues are currently under investigation.


\begin{acknowledgments}

JQ  acknowledges an Atlas Fellowship awarded by CCLRC (now STFC) in
association with St. Catherine's College, Oxford.
MH thanks the European Science Foundation (INSTANS programme) for funding a
visit to Birmingham and RAL for this work.
The authors are grateful to D.G.~Barci, E.~Fradkin, J.T.~Chalker, C.~Hooley, I.~Paul, B.J.~Powell, and
M.B.~Silva Neto for useful discussions.
JQ thanks the University of Birmingham for hospitality during the preparation of part of this manuscript.
\end{acknowledgments}



%

\end{document}